# Interface-Driven Thermo-Electric Switching Performance of VO$^+$ Diffused Soda-Lime Glass

A. Carmel Mary Esther[1*], G. Mohan Muralikrishna[1], Bonnie J. Tyler[2], Heinrich F. Arlinghaus[2], Sergiy V. Divinski[1], Gerhard Wilde[1]

[1]Institute of Materials Physics, University of Muenster, 48149-Muenster, Germany.

[2]Institute of Physics, University of Muenster, 48149-Muenster, Germany.

*Corresponding author carmelnano@gmail.com; [alphonse@wwu.de](alphonse@wwu.de)

**Graphical abstract**

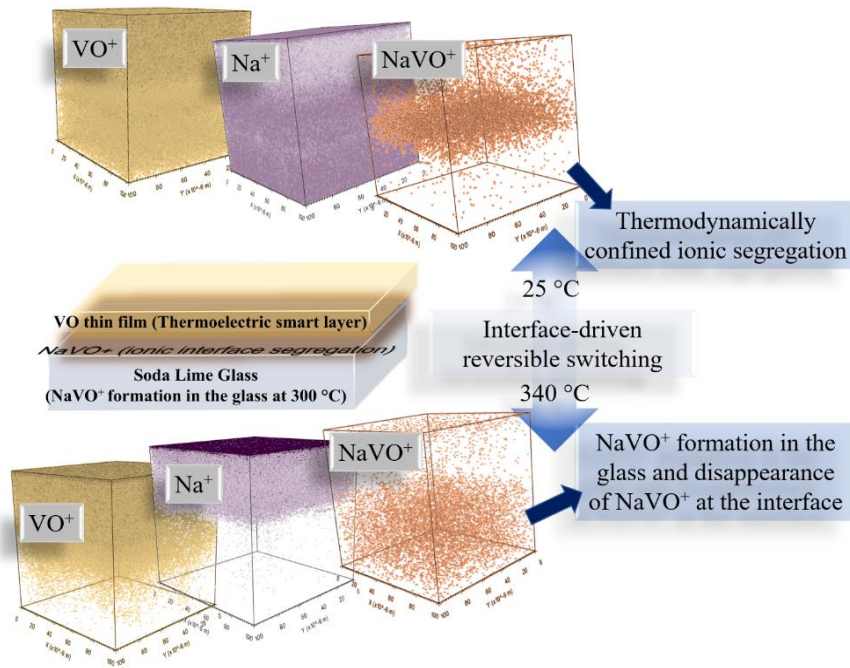

**Abstract**

Strongly confined NaVO$^+$ segregation and its thermo-responsive functionality at the interface between simple sputter-deposited amorphous vanadium oxide thin films and soda-lime glass was substantiated in the present study by in- situ temperature-controlled Time of Flight Secondary Ion Mass Spectrometry (ToF-SIMS). The obtained ToF-SIMS depth profiles provided unambiguous evidence for a reversible transformation that caused systematic switching of the NaVO$^+$/ Na$^+$ and Na$^+$/ VO$^+$ intensities upon cycling the temperature between 25 °C and 340 °C. Subsequently, NaVO complexes were found to be reversibly formed (at 300 °C) in vanadium oxide diffused glass, leading to thermo-responsive electrical behaviour of the thin film glass system. This new segregation - and diffusion-dependent multi-functionality of NaVO$^+$ points towards applications as an advanced material for thermo-optical switches, in smart windows or in thermal sensors.

## Introduction

Vanadium oxide is one of the most promising and deeply studied materials in smart and energy materials applications. Recently, the thermoelectric behaviour of vanadium oxides has been recognized as highly attractive concerning meeting energy efficiency requirements. Recent reports on the Seebeck coefficient of a vanadium pentoxide ($V_2O_5$) thin film on silica glass (−680 µV K$^{−1}$) [1] and annealed bulk $V_2O_5$ (−618 µV K$^{−1}$) [2] are highly promising for the development of a $V_2O_5$ based thermoelectric generator. Additionally, there has been an enormous focus on, and wide studies of $V_2O_5$ thin films for micro and thin-film battery applications due to their promising higher specific capacity[3]. In the attempt to replace Li-ion batteries, current research is focused mainly on Na ion-based solutions due to the abundant availability and low-cost of sodium. Thus, in addition to the studies on the Li-ion insertion and extraction mechanisms[4,5], the Na-ion insertion and extraction mechanisms for vanadium oxide-based systems is also being investigated [6,7]. The intercalation reaction of Na into the NaVO$_x$ product layer under a chemical gradient is being intensely studied for vanadium oxide-based materials in Na ion battery applications [8–10].

Although vanadium pentoxide is being explored for many functional applications, its metal-insulator transition behaviour is highly debated. Various investigations have reported significantly different temperatures at which this transition is observed, viz. ~ 127 °C[11], 260 °C[12] or 338 °C[13]. Vanadium pentoxide ($V_2O_5$) is observed to exhibit distinguishable and unique functional properties when it is deposited on different type of substrates [13–15]. Moreover, vanadium containing amorphous semiconducting glasses are also being extensively studied, for example as smart functioning glass[16]. However, there is no fundamental understanding of the mechanisms responsible for these exciting temperature dependent properties of $V_2O_5$ or their temperature dependence. For a final break-through commercialization, cost-effective and simple processing techniques must be developed along with an in-depth understanding of the material system for the entire temperature interval under consideration.

The present study focuses on the functional behaviour of cost-effectively sputter deposited vanadium oxide thin films on a commercial glass substrate such as soda-lime glass and a detailed investigation of the temperature-dependent reversible interactions of native elements in the amorphous vanadium pentoxide/soda-lime glass system by in-situ temperature-controlled Time-of-Flight Secondary-Ion Mass Spectrometry (ToF-SIMS). Additionally, temperature-dependent conductivity measurements were carried out to understand the thermo-electric behaviour of the $V_2O_5$/soda-lime glass.

**Materials and Methods**

Thin films of $V_2O_5$ were deposited on glass substrates (Thermo Fisher Scientific Silica-based microscope slides) using a vanadium metal target (99.95% purity) with a room temperature reactive sputtering process. The DC Magnetron sputtering instrument (BesTech) was operated at a power of 40 W for the current study. The vacuum recipient was pre-evacuated to a pressure $< 7.5 \times 10^{-8}$ mbar and the working pressure was adjusted to $5 \times 10^{-3}$ mbar with an O to Ar ratio of (2:3) in terms of partial pressures. The thicknesses of the V2O5 thin films were measured by a surface profilometer (Bruker DektakXT). The thicknesses were averaged over a minimum of 5 measurements on at least 2 similarly processed samples to approx. 80 nm.

The ToF-SIMS depth profiles were measured using a custom instrument which is largely equivalent to the IONTOF M6. The spectrometer was run at an operating pressure of $<10^{-8}$ mbar. The depth profiles were measured in a dual-beam mode, using a 30 keV $Bi^{3+}$ primary ion source with a pulsed current of 0.02 pA for analysis and a 1 keV Ar source with a current of 200 nA for sputtering. Analysis was performed over a $100 \times 100$ µm$^2$ area and sputtering was performed over a $500 \times 500$ µm$^2$ area. The sputtering depth was calibrated via the known film thickness and assuming a constant sputtering rate. Since we are not quantifying any diffusion properties of the elements, this approximation is suitable for the present purposes. The depth profiles are shown as dead-time corrected absolute intensities without normalization and averaged over the analysed area, assuming a lateral homogeneity of the samples over the entire area of the thin film on the glass substrate.

An Axis-Ultra spectrometer (Kratos, Manchester, UK) with charge neutralizer was used to obtain X-ray photoelectron spectra with monochromatic Al Kα radiation (hν = 1486.6 eV). Acceleration voltage and emission were set as 12 kV and 10 mA, respectively. An Impedance spectrometer (Agilent 4192A LF impedance analyser) were used to analyse the ac conductivity changes of the thin film in the temperature range 25 ˚C to 340 ˚C. For this reason, spring loaded Au contacts were configured and tested. The frequency range of 5 Hz to 2 MHz was utilized with multiple intervals. The AC voltage was set to 1.0 V for all measurements.

**Results and Discussion**

Figure 1a show the depth profile and mass spectrum (See Extended Data Figure 1) of soda-lime glass, respectively, substantiating the presence of $Na^+$ as a major constituent of the glass along with minor $K^+$ and $Ca^+$ additives. The intensities of $Na^+$ are found to be almost constant through the sputtered glass substrate.

The vanadium oxide thin film was sputter deposited on soda-lime glass as explained in the Materials and Methods section. Figure 1b shows the ToF-SIMS depth profile through $V_2O_5$ analysed at 25 °C. The signals for $VO^+$, $Na^+$, $NaVO^+$ and $Si^+$ are shown. The ToF-SIMS spectrum also contains signals from $V^+$ and a wide range of $V_xO_y^+$ cluster ions in the $V_2O_5$ layer as well as traces of $K^+$, $Ca^+$ and $Al^+$ in the glass substrate. These signals are not included in Figure 1b for the sake of clarity.

The ToF-SIMS depth profile shows the presence of $Na^+$ through the entire thickness (~80 nm) of the thin film already in the as-deposited state. This is likely due to diffusion of Na into the amorphous vanadium oxide thin film during the sputter deposition with the soda-lime glass substrate acting as a source for $Na^+$. The surface is found to be enriched with oxygen, suggesting the presence of higher oxide states of vanadium along with minute trace amounts of $Na^+$. Further, the intensity of $Na^+$ is negligible at the surface, which is also supported by XPS analysis. Extended Figure 2 a and b shows results of X-ray Photoelectron Spectroscopy (XPS) on a vanadium oxide thin film. The survey spectrum (See Extended Data Figure 2) clearly reveals the presence of V and O species in the deposited thin films along with low levels of C and N, typical of organic contaminants commonly observed on samples that have been exposed to ambient air[17]. No traces of Na were revealed on the XPS survey spectra as shown in Fig. S2 a. The interface between the thin film and the glass substrate is characterized by enrichment/segregation of $Na^+$ and the formation of $NaVO^+$ (see Figure 1). The $NaVO^+$ profile is roughly Gaussian shaped and an inverse Gaussian-like peak is observed for $V^+$ at the film/substrate interface. $Na^+$ diffusion is not expected to alter the oxidation state of V by the negative charge-feedback mechanism for the transition metal oxides as explained by Raebiger *et al.* [18]

Extended Figure 2b shows the core level XPS spectrum of V and O. The binding energies of 517.3 and 530.2 eV were detected for $V2p_{3/2}$ and $O1s$, respectively, in the as-prepared thin film $V_2O_5$ / glass. The binding energy splitting between the $V2p_{3/2}$ and $V2p_{1/2}$ states was found to be 7.2 eV. The peak values are in good agreement with the previously reported data for stoichiometric $V_2O_5$ at room temperature within an experimental uncertainty of about ±0.1eV[13]. Since the core level peaks of $O1s$ and $V2p_{3/2}$ are narrow and represent a single oxidation state, the peaks were not deconvoluted further to quantify minor oxidation states.

Interestingly, the ToF-SIMS depth profile showed unambiguous evidence for the occurrence of (sputtering-induced) diffusion of $V^+$ and its oxides into the glass substrate. The ToF-SIMS mass spectrum of soda-lime glass shown in Extended Figure 1 substantiates the absence of V in the substrate prior to the film deposition. Hence, diffusion of $V^+$ into the glass and the formation of $NaVO^+$ near/at the interface occurs either during or after the sputter deposition of the $V_2O_5$ thin film.

Diffusion of $Na^+$ from silica-based glass into a vanadium oxide thin film has been reported previously [19]. On the other hand, soda-lime glass as a substrate for $VO_2$ coating was very rarely utilized[20]. It was found to be highly challenging to optimize the coating procedure for thermochromic applications[21]. However, to the best of our knowledge, there are no reports available demonstrating segregation formation of $NaVO^+$ at the film/substrate interface. An order-of-magnitude estimate of the sputter-induced diffusion coefficient of $Na^+$ in silica suggest a value of about $3\times10^{-20}$ m$^2$/s which would correspond to the value predicted for Na diffusion in $SiO_2$ at 115 °C or 107 °C according to the tracer measurements of Frischat[22] or Tanguep Njiokep & Mehrer [23], respectively. The estimated effective temperatures are found to be much lower than the experimentally predicted activation temperature i.e. less than 200 °C required to initiate Na diffusion into the thin film[21].

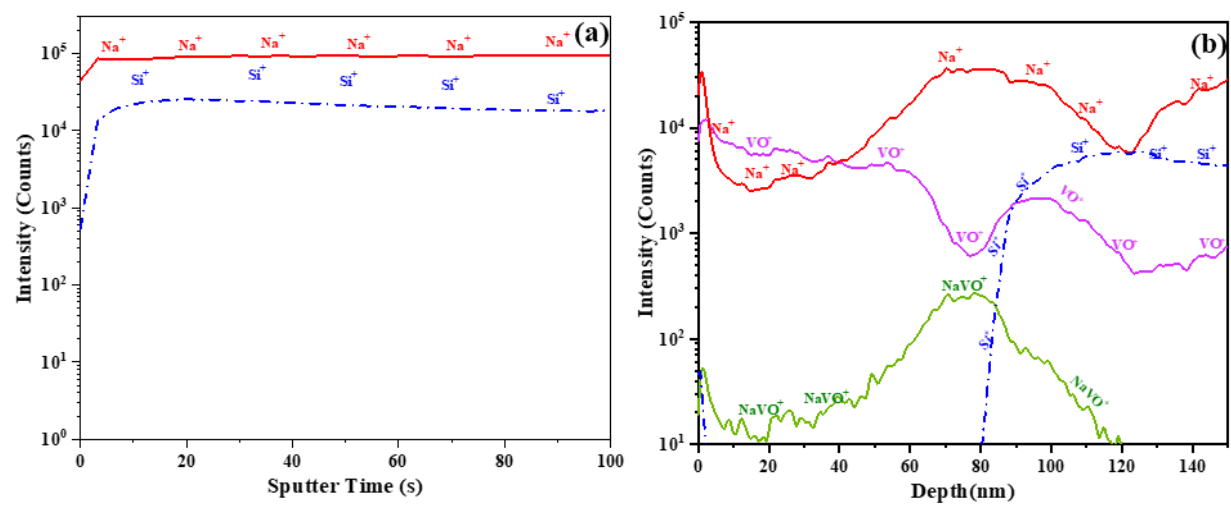

**Figure 1**. ToF-SIMS intensity profiles of (a) pristine glass as a function of sputter time and (b) $V_2O_5$ on glass substrate as a function of depth

In order to investigate the reversible semiconductor-to-metal transition (SMT) behaviour of amorphous $V_2O_5$, the thin film system was heat-treated in the temperature range of 25-340 ˚C and the ToF-SIMS analysis was (concurrently) performed in-situ.

For the soda-lime glass, the ToF-SIMS profile obtained at 300 ˚C showed no change of the element intensities except a minor depletion of $Na^+$ near the surface (see Extended Data Figure 3). The complete set of the depth profiles of $V_2O_5$ / glass is shown in Extended Data Figure 4 for all temperatures under investigation. Strikingly, $NaVO^+$ appeared prominently at 300 ˚C in the vanadium-diffused glass and the intensity of the $NaVO^+$ signal remained constant up to 340 ˚C.

In Figure 2, the ToF-SIMS depth profiles recorded for $VO^+$, $Na^+$ and $NaVO^+$ during thermal cycling are shown for two limiting temperatures of 25 °C and 340 ˚C. An enrichment of $Na^+$ ions is prominent at the $V_2O_5$ / glass interface at temperatures above 300 ˚C (See Extended Data Figure 4). During heating, the intensity of V and its oxides is seen to increase, their distribution is flattened out and the sputter-induced "gap" at the interface disappears upon heating to 340 ˚C. The intensity of the $NaVO^+$ signal is also reduced at the interface at 340 ˚C as shown in Fig. 2 (c) and it appears again when the temperature is decreased to 25 °C. Simultaneously, an interface-related gap in the $VO^+$ distribution is formed when the temperature is lowered below 300 °C.

Extended Data Figure 4 substantiates a systematic and reversible evolution of the $Na^+$, $NaVO^+$ and $VO^+$ signals. Remarkably, the initial (as-sputtered) intensity levels are retained upon cooling to 25 ˚C. Simultaneously, the $NaVO^+$ interface and the substrate-related peaks disappear. The sample was cycled three times between 25 ˚C and 340 ˚C to ensure the reversibility of the aforementioned reactions presented in Fig. 3 (a-c).

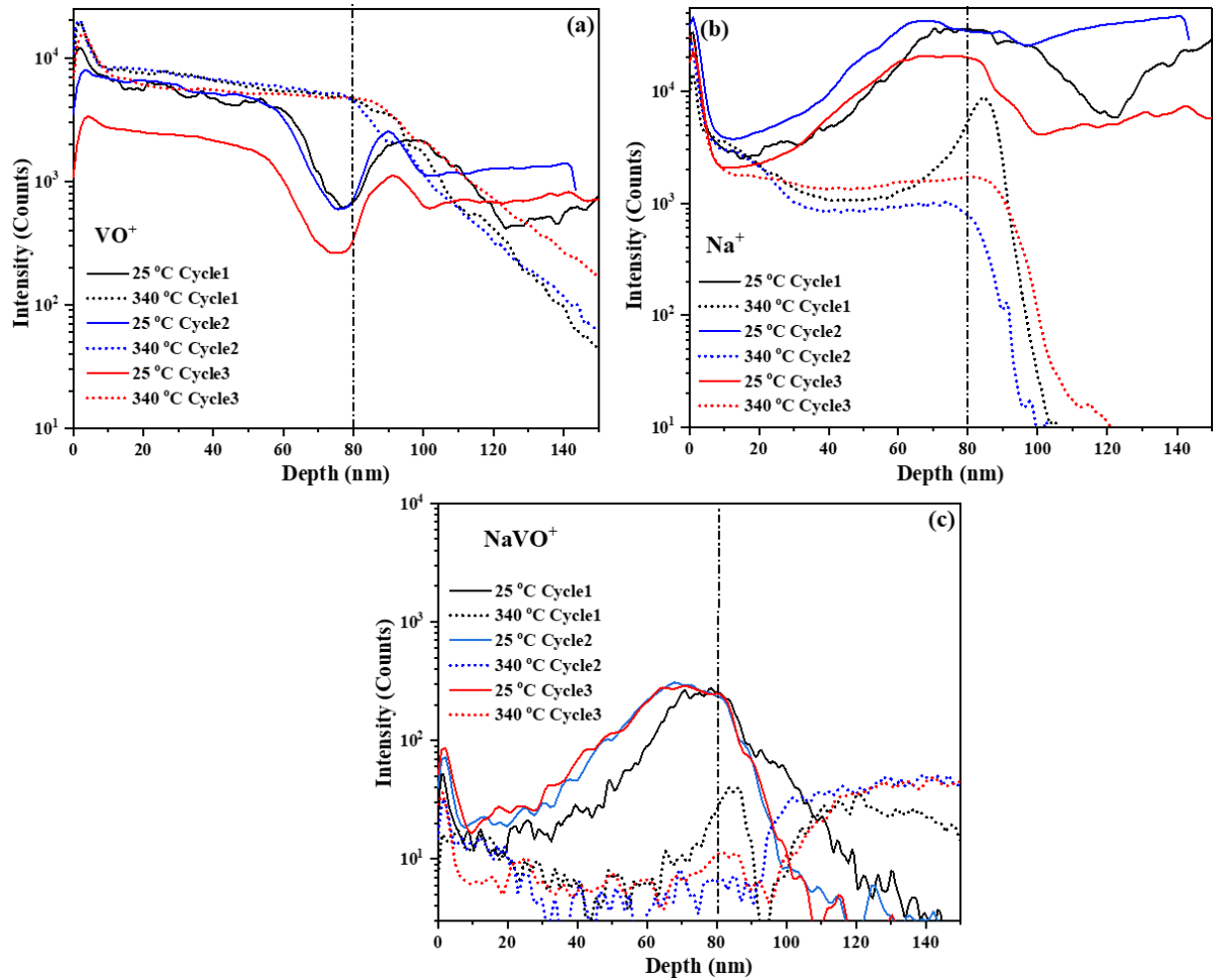

**Figure 2**. In-situ ToF-SIMS depth profiles of $V_2O_5$ on glass, comparing three subsequent temperature cycles between 25 ˚C and 340 ˚C and analysing the distribution of a) $VO^+$, b) $Na^+$ and c) $NaVO^+$

Figure 3 shows the 3D maps corresponding to the distributions of $VO^+$, $Na^+$ and $NaVO^+$ at 25 ˚C and 340 ˚C, respectively. It is evident that the expected reaction Na + VO ⇔ NaVO occurs uniformly in planes parallel to the film/substrate interface for the whole ToF-SIMS sputtered area of 100×100 µm$^2$.

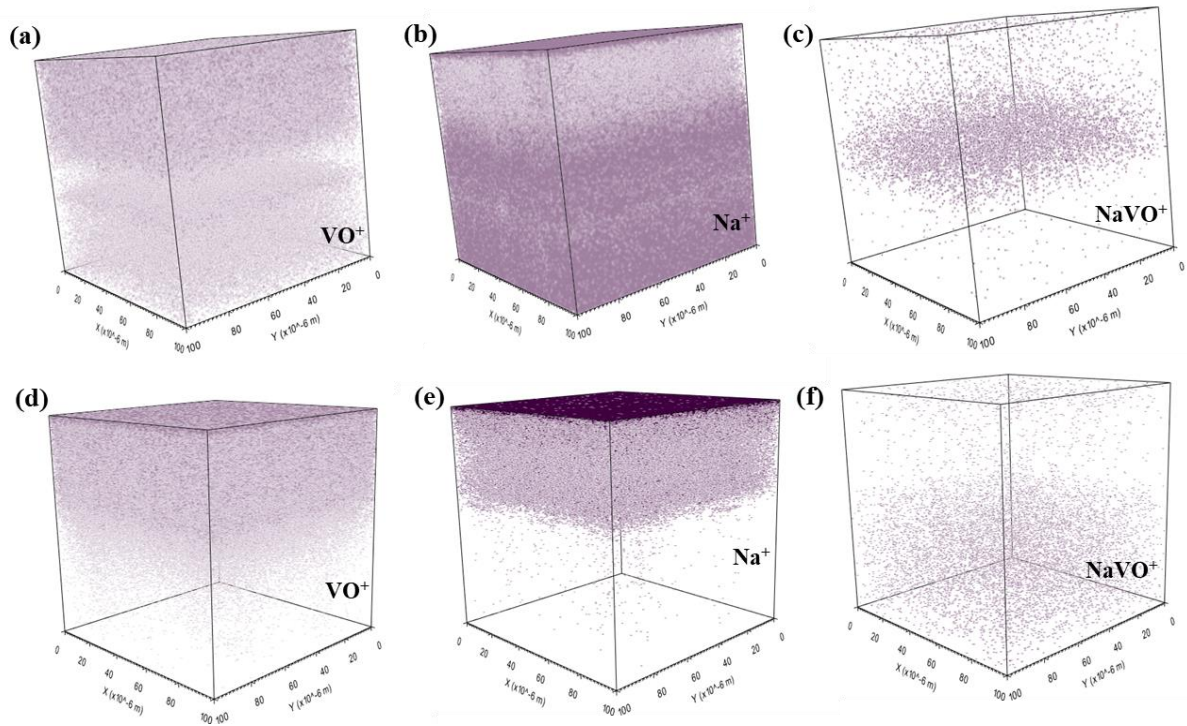

**Figure 3**. 3D maps of the distributions of $VO^+$, $Na^+$ and $NaVO^+$ at (a, b, c) 25 °C and (d, e, f) 340 °C, respectively

Figure 4 substantiates a switching behaviour (complete reversibility) of the relative intensity of $NaVO^+$ with respect to the total intensities of $VO^+$ and $Na^+$ during all thermal cycles. This revealed that the major changes occurred at the interface and in the substrate but not at the surface of the thin film. In other words, the stable surface oxide state of $V_2O_5$ helped as a protective/passivation layer obstructing secondary reactions. The observed reactions are reproducible and unambiguous as all thermal cycles for the in-situ ToF-SIMS experiments were conducted under high vacuum conditions.

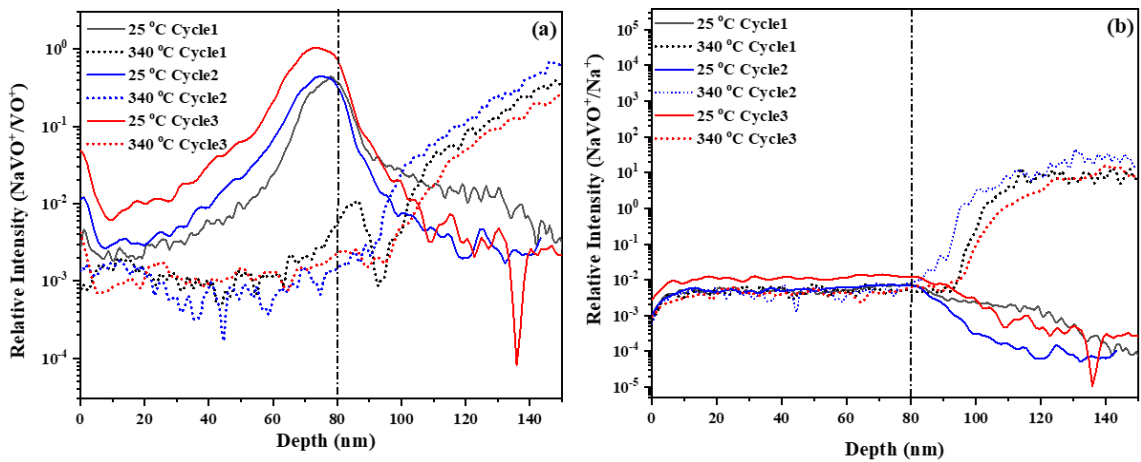

**Figure 4**. Relative intensity ratios of (a) $NaVO^+ / VO^+$ and (b) $NaVO^+ / Na^+$ comparing three subsequent temperature cycles between 25 °C and 340 °C

Figure 5a and b show the first and third cycles of the conductivity measurements as a function of both frequency and temperature of the $V_2O_5$/soda-lime glass, respectively. The initial surface conductivity level of as-deposited amorphous $V_2O_5$/soda-lime glass is much lower (8.23 x $10^{-10}$ S/cm). As shown in Figure 5a, the low-frequency response of the conductivity substantiates that the initial stabilization of the thin film/glass system occurred below 200 ˚C. The stabilization might be due to the dissociation of adsorbed oxides and water molecules from the surface of the thin film. After the stabilization of the initial reactions, the conductivity varies reproducibly between the values of ~2 x $10^{-8}$ S/cm and ~4 x $10^{-7}$ S/cm for 25 ˚C and 340 ˚C, respectively, for all subsequent cycles. The experimental behaviour is well matched with the literature reports for the bulk vanadium-based semiconducting glass system [24–27], except for the particular features of the high-frequency (5 – 2 MHz) response. Gradual and reversible changes of the response between 300 ˚C and 340 ˚C at these frequencies are observed. The formation of $NaVO^+$ is prominent just at 300 ˚C and its fraction remains constant in the glass, which supports the assumption of a linear ac conductivity behaviour of the glass system. Hence, in the present case the observed changes in the dielectric component of the $V_2O_5$/soda-lime glass strongly correlate with the reversible, segregation-induced interfacial production/decomposition of $NaVO_x$ alone.

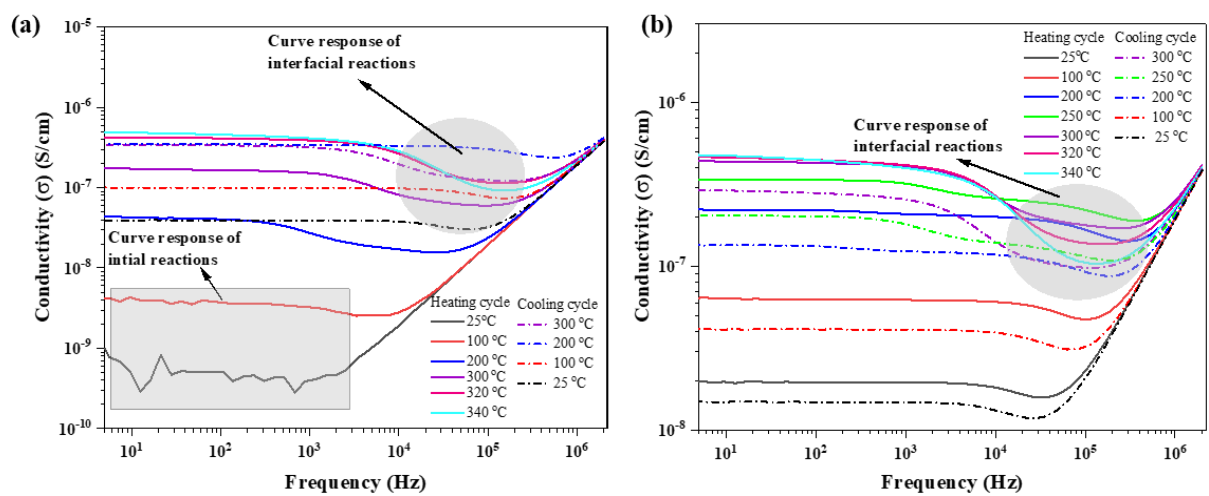

**Figure 5**. Frequency response of conductivity of the $V_2O_5$ thin film on glass for the (a) first and (b) third cycle of the thin film three subsequent temperature cycles between 25 ˚C and 340 ˚C

## Advanced Functionalities

Three unique temperature-dependent reaction mechanisms are found in the $V_2O_5$/soda-lime glass system.

(i) Change of electrical conductivity of the $V_2O_5$ / glass system

In the present study, the temperature dependent conductivity increments (Figure 5 a, b) and the ToF-SIMS results elucidating reversible temperature-dependent solid-state reactions in the $V_2O_5$ thin film /soda-lime glass provide evidence of the possibility for the fabrication of a thermoelectric generator[16,28] or smart window applications[29], as have been preliminarily attempted by others for vanadium oxides and its composites[30–35]. Further, here it is important to emphasise that the investigated $V_2O_5$/soda-lime glass system displays at least 80% optical transparency in the visible spectral region with an optical band gap of 2.3eV. This property combination is highly beneficial for the fabrication of transparent thin-film based thermo-electric generators and thermal sensor devices.

(ii) Formation of $NaVO^+$ in the glass matrix

Vanadium, with its stabilized oxide and ionic states, diffused into the glass matrix in the presence of $Na^+$. During heating, $Na^+$ and $VO^+$ combined and formed $NaVO^+$ (Supplementary details Figure S2). Ultimately, the reaction of Na + VO = NaVO is expected to produce excess electrons. Upon cooling, the $NaVO^+$ separated into ionic states of $Na^+$ and $VO^+$, which is already demonstrated as stored charge concentration in the dielectric glass medium. The current study verifies charging and discharging mechanisms by solid state reactions upon thermal cycling. Hence the current results indicate that the produced electricity from the thermal behaviour of vanadium oxides can be stored directly into the soda-lime glass.

(iii) Thermal response of the confined ionic interface layer

The systematic alteration of ionic states (VO and NaVO complex) by segregation shown by in-situ ToF-SIMS (Supplementary details Figure S2) were well correlated with the concurrently recorded higher frequency response of the electrical conductivity (Fig. 5). Since we expect the $NaVO^+$ nano-interface to outperform with conventional thin film-based devices regarding the electrical conductivity and optical response, we are proposing a conceptual design of a 'thermodynamically confined ionic interface/segregation layer' to serve as a thermo-electrical/thermo-optical switch or sensor.

## Concluding Remarks

The present work primarily focused on the meticulous characterization of a vanadium oxide thin film coated on commercial glass to reveal its smart functional behaviour. The observed results substantiate the formation of a thermodynamically confined ionic interface. The diffused vanadium in the glass was introduced by a simple sputter deposition process. Furthermore, the thermal behaviour of the thin film/ glass system is highly reversible and stable. Recent in-situ TEM work [36] illustrates that no bulk transitions, such as an oxidation state change or the crystallization of the amorphous vanadium pentoxide thin film, occur below 400 ˚C. Permanent vanadium oxidation state intensity changes were not observed in the depth profiles obtained by ToF-SIMS.

For the first time, the production of interface-confined $NaVO^+$ via a solid-state thermal reaction of vanadium pentoxide with sodium ions supplied by the soda-lime glass substrate is documented. The observed temperature dependent conductivity increments commonly reported for vanadium oxide based and vanadium containing glass could be explained by thermal activation in semiconducting glasses [24–27]. However, these results provide evidence for metal- metal interactions of vanadium oxide during SMT, which induced the conjugation of $VO^+$ with $Na^+$ to form $NaVO^+$ in the glass.

Characterization of this $V_2O_5$/soda-lime glass remains challenging and opens new research directions as prospects for future work:

- Depth dependent oxidation state and valence state determination of vanadium
- Diffusion kinetics of vanadium in the complex glass in the presence of $Na^+$
- Interface formation mechanism by different processing methods and conditions and the actual conditions responsible for the observed thermal behaviour of the interface
- The importance of soda-lime glass for the formation of NaVO
- Determination of electrical and optical constants of individual layer systems (thin film, interface and $V^+$ diffused glass) as a function of temperature, and
- Designing electrical components of $V_2O_5$/soda-lime glass

In conclusion, the present study demonstrates the potential of $V_2O_5$ thin films for a self-chargeable transparent thin film thermo-electric generator on a commercial window glass and many more applications which can be realised in the near future. The obtained results of vanadium diffusion in the glass show new perspectives concerning the basic understanding of the thermal response of vanadium oxide-based semiconducting glass bulk systems.

**Acknowledgments**

Authors are grateful to Manfred Bartsch, Physics Institute, WWU, Germany for help in conducting the XPS of samples for the current study. ACME would like to acknowledge AvH Foundation, Germany for a Post-Doctoral fellowship award to conduct research at WWU, Germany. GMM would like to acknowledge DAAD, Germany for awarding a Fellowship to conduct experiments at WWU. SVD and GW would like to acknowledge the German Science Foundation (DFG) for partial financial support.

**References**

1. Bianchi, C. *et al.* V2O5 Thin Films for Flexible and High Sensitivity Transparent Temperature Sensor. *Adv. Mater. Technol.* **1**, 1600077 (2016).

2. Yang, Z., Bin Jia, X., Nkemeni Darrin, S., Hui Li, G. & Min Zhou, S. Pure V2O5 high-electricity transmission properties at low temperatures by structural changes. *Mater. Lett.* **279**, 128493 (2020).

3. Liu, P. C., Zhu, K. J., Gao, Y. F., Luo, H. G. & Lu, L. Recent progress in the applications of vanadium-based oxides on energy storage: From low-dimensional nanomaterials synthesis to 3D micro/nano-structures and free-standing electrodes fabrication. *Adv. Energy Mater.* **7**, 1700547 (2017).

4. Ma, W. *et al.* Impacts of Surface Energy on Lithium Ion Intercalation Properties of V2O5. *ACS Appl. Mater. Interfaces* **8**, 19542–19549 (2016).

5. Pan, A. *et al.* Facile synthesized nanorod structured vanadium pentoxide for high-rate lithium batteries. *J. Mater. Chem.* **20**, 9193–9199 (2010).

6. Song, W. *et al.* Exploration of ion migration mechanism and diffusion capability for Na 3V2(PO4)2F3 cathode utilized in rechargeable sodium-ion batteries. *J. Power Sources* **256**, 258–263 (2014).

7. Song, W. *et al.* First exploration of Na-ion migration pathways in the NASICON structure Na3V2(PO4)3. *J. Mater. Chem. A* **2**, 5358–5362 (2014).

8. Guignard, M. *et al.* P2-Na x VO 2 system as electrodes for batteries and electron-correlated materials. *Nat. Mater.* **12**, 74–80 (2013).

9. Tepavcevic, S. *et al.* Nanostructured bilayered vanadium oxide electrodes for rechargeable sodium-ion batteries. *ACS Nano* **6**, 530–538 (2012).

10. Kim, S.-W., Seo, D.-H., Ma, X., Ceder, G. & Kang, K. Electrode Materials for Rechargeable Sodium-Ion Batteries: Potential Alternatives to Current Lithium-Ion Batteries. *Adv. Energy Mater.* **2**, 710–721 (2012).

11. Blum, R. P. *et al.* Surface metal-insulator transition on a vanadium pentoxide (001) single crystal. *Phys. Rev. Lett.* **99**, 3–6 (2007).

12. Kang, M., Kim, I., Kim, S. W., Ryu, J. W. & Park, H. Y. Metal-insulator transition without structural phase transition in V 2 O5 film. *Appl. Phys. Lett.* **98**, 2009–2012 (2011).

13. Porwal, D. *et al.* Study of the structural, thermal, optical, electrical and nanomechanical properties of sputtered vanadium oxide smart thin films. *RSC Adv.* **5**, 35737–35745 (2015).

14. Prajwal, K., Carmel Mary Esther, A. & Dey, A. RF transparent vanadium oxide based single and bi-layer thin films as passive thermal control element for satellite antenna application. *Ceram. Int.* **44**, 16088–16091 (2018).

15. Esther, A. C. M., Dey, A., Rangappa, D. & Sharma, A. K. Low reflectance sputtered vanadium oxide thin films on silicon. *Infrared Phys. Technol.* **77**, 35–39 (2016).

16. Hassaan, M. Y., Osman, H. M., Hassan, H. H., El-Deeb, A. S. & Helal, M. A. Optical and electrical studies of borosilicate glass containing vanadium and cobalt ions for smart windows applications. *Ceram. Int.* **43**, 1795–1801 (2017).

17. Lu, Y., Liu, L., Mandler, D. & Lee, P. S. High switching speed and coloration efficiency of titanium-doped vanadium oxide thin film electrochromic devices. *J. Mater. Chem. C* **1**, 7380–7386 (2013).

18. Raebiger, H., Lany, S. & Zunger, A. Charge self-regulation upon changing the oxidation state of transition metals in insulators. *Nature* **453**, 763–766 (2008).

19. Alamarguy, D., Castle, J. E., Liberatore, M. & Decker, F. Distribution of intercalated lithium in v2O5 thin films determined by SIMS depth profiling. in *Surface and Interface Analysis* vol. 38 847–850 (John Wiley & Sons, Ltd, 2006).

20. Wang, X. *et al.* Preparation of thermochromic VO2 thin films on fused silica and soda-lime glass by RF magnetron sputtering. *Japanese J. Appl. Physics, Part 1 Regul. Pap. Short Notes Rev. Pap.* **41**, 312–313 (2002).

21. Miller, M. J. & Wang, J. Influence of Na diffusion on thermochromism of vanadium oxide films and suppression through mixed-alkali effect. *Mater. Sci. Eng. B Solid-State Mater. Adv. Technol.* **200**, 50–58 (2015).

22. FRISCHAT, G. H. Sodium Diffusion in SiO2 Glass. *J. Am. Ceram. Soc.* **51**, 528–530 (1968).

23. Tanguep Njiokep, E. M. & Mehrer, H. Diffusion of 22Na and 45Ca and ionic conduction in two standard soda-lime glasses. *Solid State Ionics* **177**, 2839–2844 (2006).

24. Barczyński, R. J., Król, P. & Murawski, L. Ac and dc conductivities in V2O5-P2O 5 glasses containing alkaline ions. in *Journal of Non-Crystalline Solids* vol. 356 1965–1967 (North-Holland, 2010).

25. Barde, R. V., Nemade, K. R. & Waghuley, S. A. AC conductivity and dielectric relaxation in V 2 O 5 -P 2 O 5 -B 2 O 3 glasses. *J. Asian Ceram. Soc.* **3**, 116–122 (2015).

26. Murawski, L. & Barczyński, R. J. Electronic and ionic relaxations in oxide glasses. *Solid State Ionics* **176**, 2145–2151 (2005).

27. Murawski, L. A.c. conductivity in binary V2O5-P2O5 glasses. *Philos. Mag. B Phys. Condens. Matter; Stat. Mech. Electron. Opt. Magn. Prop.* **50**, 69–74 (1984).

28. Yang, Y. *et al.* Transparent lithium-ion batteries. *Proc. Natl. Acad. Sci. U. S. A.* **108**, 13013–13018 (2011).

29. Long, L. & Ye, H. How to be smart and energy efficient: A general discussion on thermochromic windows. *Sci. Rep.* **4**, 1–10 (2014).

30. Behera, M. K., Williams, L. C., Pradhan, S. K. & Bahoura, M. Reduced Transition Temperature in Al:ZnO/VO2 Based Multi-Layered Device for low Powered Smart Window Application. *Sci. Rep.* **10**, 1–11 (2020).

31. Zhou, J. *et al.* VO 2 thermochromic smart window for energy savings and generation. *Sci. Rep.* **3**, 1–5 (2013).

32. Zhang, J. *et al.* Hydrothermal growth of VO2 nanoplate thermochromic films on glass with high visible transmittance. *Sci. Rep.* **6**, (2016).

33. Shao, Z., Cao, X., Luo, H. & Jin, P. Recent progress in the phase-transition mechanism and modulation of vanadium dioxide materials. *NPG Asia Mater.* **10**, 581–605 (2018).

34. Liu, S. *et al.* Bio-inspired TiO2 nano-cone antireflection layer for the optical performance improvement of VO2 thermochromic smart windows. *Sci. Rep.* **10**, 1–14 (2020).

35. Dey, A. *et al.* Nanocolumnar Crystalline Vanadium Oxide-Molybdenum Oxide Antireflective Smart Thin Films with Superior Nanomechanical Properties. *Sci. Rep.* **6**, (2016).

36. Alphonse, C. M. E., Garlapati, M. M., Hilke, S. & Wilde, G. Double shadow masking sample preparation method for in-situ TEM characterization. *Nano Sel.* nano.202000063 (2020) doi:10.1002/nano.202000063.

**Extended Data Figures**

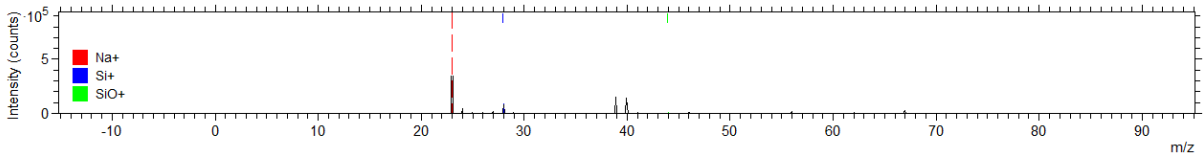

**Figure 1.** Mass spectrum of uncoated soda-lime glass which shows the absence of vanadium and its oxides

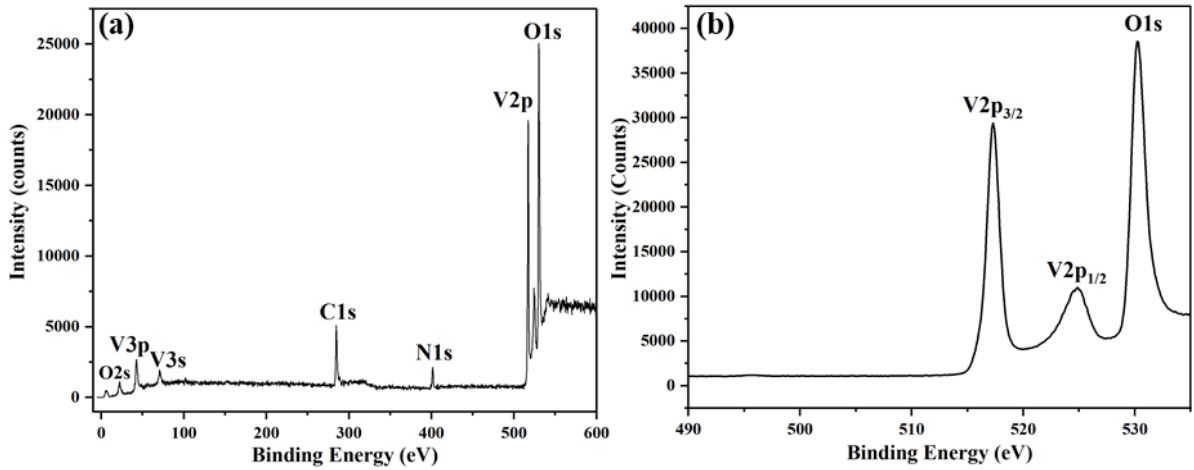

**Figure 2**. XPS of $V_2O_5$ on soda-lime glass: (a) survey spectrum and (b) core level spectrum

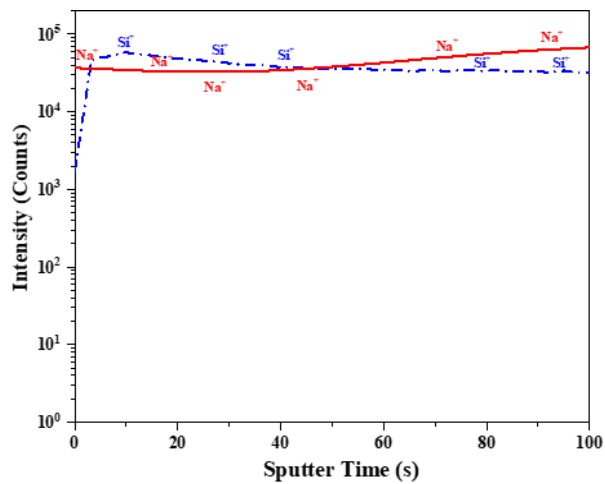

**Figure 3.** SIMS profile of Bare Glass at 300 ˚C

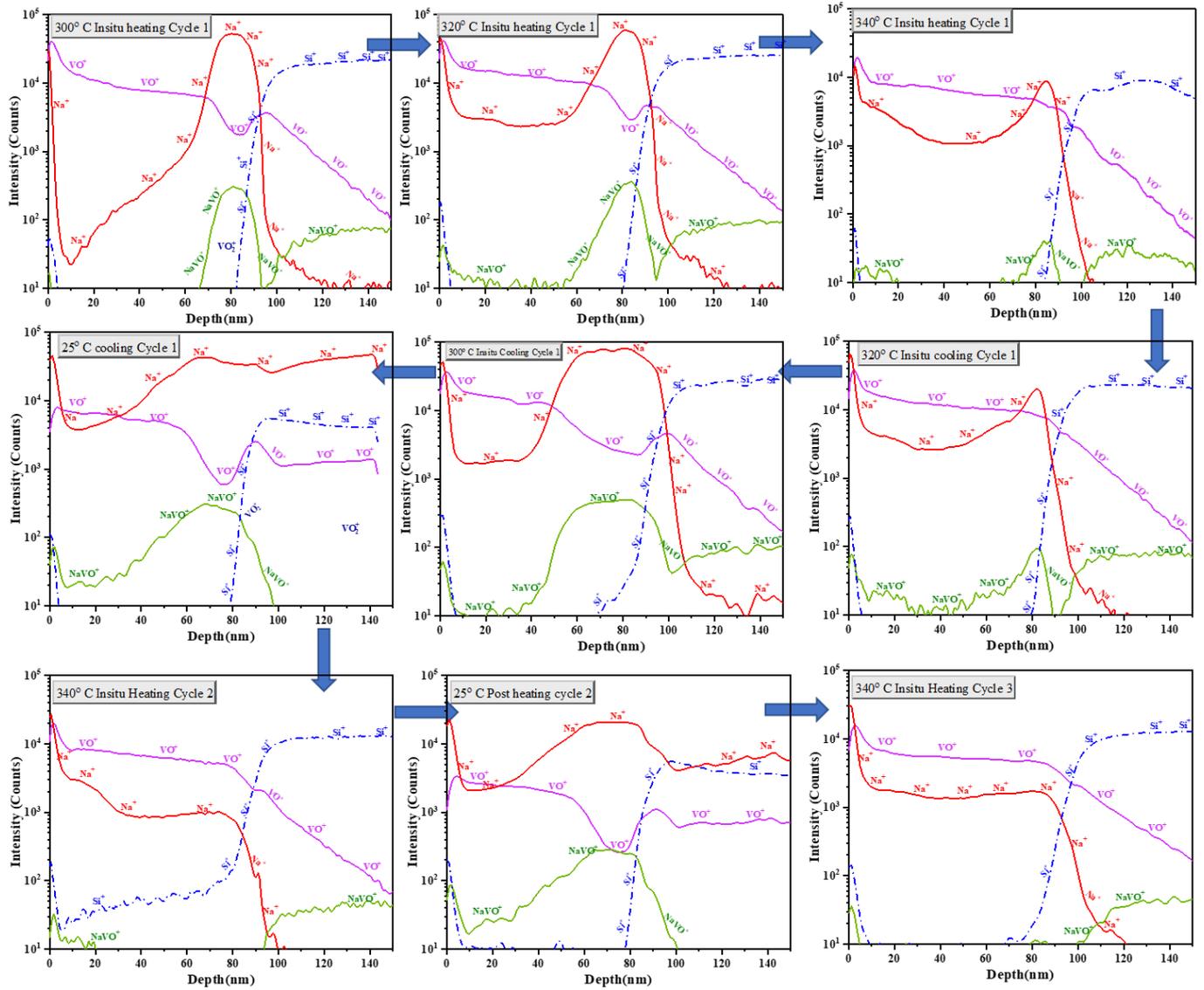

**Figure 4.** In-situ temperature based ToF-SIMS depth profiles cycles of $V_2O_5$/glass. Time sequence showed by arrows starting from 300 °C to 340 °C and subsequent cooling cycle followed by cycles 2 and 3.